\documentclass[12pt,a4paper]{article}
\usepackage{geometry}
\usepackage[utf8]{inputenc}
\usepackage[english]{babel}
\usepackage[T1]{fontenc}
\usepackage{natbib}
\usepackage{graphicx}

\defcitealias{Ercolano2010}{ECH11}

 \title{ \begin{flushleft}\bf The lifetime of protoplanetary discs: \newline
Observations and Theory \end{flushleft}}
 \author{Barbara Ercolano and Christine Koepferl}
 \date{}
 
\begin{document}
\maketitle

\begin{abstract}
The time-scale over which and modality by which young stellar objects
(YSOs) disperse their circumstellar discs dramatically influences the eventual formation
and evolution of planetary systems. By means of extensive radiative transfer
(RT) modelling, we have developed a new set of diagnostic diagrams in the infrared
colour-colour plane (K\,-\,[24] vs. K\,-\,[8]), to aid with the classiffication of the
evolutionary stage of YSOs from photometric observations. Our diagrams allow the
differentiation of sources with un-evolved (primordial) discs from those evolving
according to different clearing scenarios (e.\,g.\,homologous depletion vs. inside-out
dispersal), as well as from sources that have already lost their disc. Classification of
over 1500 sources in 15 nearby star-forming regions reveals that approximately 39\,\% of the sources lie in the primordial disc region, whereas between 31\,\% and 32\,\% disperse from the inside-out and up to 22\,\% of the sources have already lost their disc.
Less than 2\,\% of the objects in our sample lie in the homogeneous draining regime.
Time-scales for the transition phase are estimated to be typically a few $10^{5}$ years
independent of stellar mass. Therefore, regardless of spectral type, we conclude that
currently available infrared photometric surveys point to fast (of order 10\,\% of the
global disc lifetime) inside-out clearing as the preferred mode of disc dispersal.
\end{abstract} 

\begin{minipage}{21cm}
\begin{tabular}{l}
\hline
\hline
\\
{\footnotesize\bf Barbara Ercolano}\\
\footnotesize Excellence Cluster ‘Universe’, Boltzmannstra\ss e 2, 85748 Garching, Germany \\
\footnotesize e-mail: ercolano@usm.lmu.de\\
{\footnotesize\bf Christine Koepferl}\\
\footnotesize Universit\"ats-Sternwarte M\"unchen, Scheinerstra\ss e 1, 81679 M\"unchen, Germany \\
\footnotesize e-mail: koepferl@usm.lmu.de
\end{tabular}
\end{minipage}

\section{Introduction}
The lifetime and modality for the dispersal of protoplanetary discs around newly
formed low mass stars (approximately solar mass or lower) is a key parameter that
influences the formation and evolution of eventual planetary systems. The classical
picture that emerged from the last decade of photometric observations, mainly carried
out with the Spitzer Space telescope is that of disc evolution being described by
two different timescales. The first timescale could be defined as a ’global timescale’,
i.\,e.\,the total time it takes from a star to go from disc-bearing to disc-less, and a ’dispersal
timescale’, i.\,e.\,the time it takes for a disc to disappear once dispersal has
set in. Global timescales, which can be inferred from the study of disc frequencies
(e.\,g.\,\citet{Haisch2001}), are of order a few million years (e.\,g.\,\citet{Hernandez2007b, Mamajek2009}). Dispersal time-scales, as determined from the study of infrared colours of young stars, appear to be much shorter, indicating that the dispersal mechanism
must be fast (e.\,g.\,\citet{Kenyon1995}; \citet{Luhman2010}; \citet{Ercolano2010}). Such observed two-timescale behaviour has favoured the development
of disc dispersal models that involve a rapid disc clearing phase, contrary to
the predictions of simple viscous draining, and in agreement with photoevaporation
\citep{Clarke2001, Alexander2006a, Alexander2006b, Ercolano2008, Ercolano2009, Gorti2009, Owen2010, Owen2011b, Owen2011a, Owen2012a} or possibly
planet formation \citep{Armitage1999}.
The interpretation of infrared colours in relation to the evolutionary state of a
disc is, however, far from being trivial. This is particularly true with regards to the
classification of transition discs, the latter being intended as objects caught in the act
of disc dispersal. The evolution of the dust component in a disc is mirrored by the
evolution of colours in the infrared plane. By means of radiative transfer modelling,
 \citeauthor*{Ercolano2010} (2011, henceforth \citetalias{Ercolano2010}) identified the regions in the K\,-\,[8] vs. K\,-\,[24] plane where primordial discs, discs with inner-holes (i.\,e.\,presumably
being dispersed from the inside-out) and discs which lose mass homogeneously at all
radii, are expected to be found. Their study, which was limited to M-stars, showed
that in the case of the cluster IC348, most discs disperse from the inside-out and undergo
the transition on a short time-scale, as predicted by standard photoionisation
models. These conclusions are in contrast with the conclusions of \citet{Currie2009b}, who claimed instead a large number of ’homogeneously depleting’ discs,
for the same cluster. Such discrepancies highlight the need for detailed modelling
in the interpretation of IR colours of discs. The study of \citetalias{Ercolano2010} was restricted to
M-stars in only one cluster, which prevented the authors from being able to make a
more general statement with regards to disc dispersal. Here we present results from
a forthcoming paper \citep{Koepferl2012}, which significantly improves on the work
of \citetalias{Ercolano2010} by performing further RT calculation to evaluate evolutionary tracks in the IR colour plane for stars of different spectral types. We then apply our results to the photometric data of 15 nearby star-forming regions, that we collected from
the literature, in order to address the question of what is the preferred mode of disc
dispersal.
\section{Disc evolution in the Infrared two colour plane}
\begin{figure}[h]
\centering
\begin{minipage}[b]{0.55\textwidth}
\vspace{0pt}
\includegraphics[width=\textwidth]{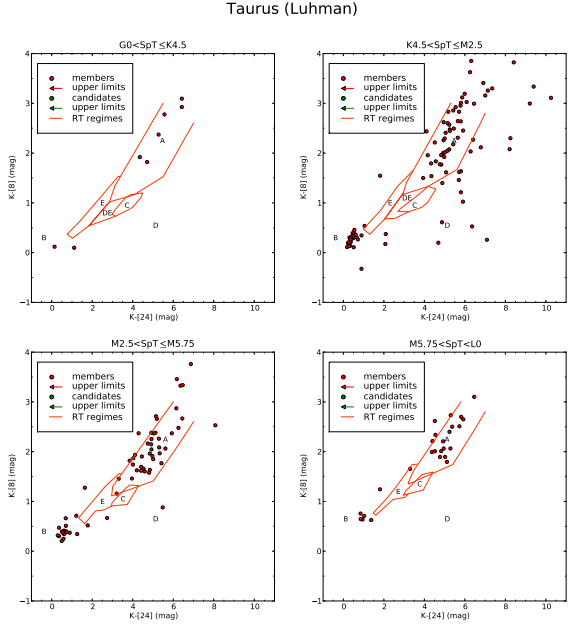} 
\end{minipage}%
\hfill
\begin{minipage}[b]{0.45\textwidth}
\caption{Disc evolution diagnostic diagram applied to the YSOs in the 1 Myr old cluster Taurus. 20 sample points lie outside the limits of this plot, but they are still included in the final statistics. (51\,\% primordial optically thick (A), 0\,\% primordial ultra-settled (C), 23\,\% disc-less (B), 14\,\% inside-out clearing (D), $<1$\,\% homogeneous draining (E)).}
\label{Luhman}
\label{time}
\end{minipage}
\end{figure}

By means of radiative transfer modelling we identify 5 areas in the K\,-\,[8] vs. K\,-\,[24] plane corresponding to discs at different evolutionary stages or with different
geometries, these are shown in Figure~\ref{Luhman}. Namely, primorial optically thick (flared and/\,or mixed) discs occupy
area A, while disc-less objects cluster in area B and area C is for primorial ultra-settled (flat) discs. Discs that are in the act of dispersal belong to area D and E, where the former
is for discs that are clearing from the inside-out, and the latter is for discs that are
progressively going optically thin homogeneously at all discs radii (as would be
expected from viscous evolution alone).
Our classification scheme allows us to use currently available infrared photometric
surveys from nearby star-forming regions in order to address the question of what
is the preferential mode of disc dispersal.We therefore applied our colour-colour diagnostic
diagrams to classify 1529 objects in 15 nearby star-forming regions. As an
example, we overplot the data for Taurus (taken from \cite{Luhman2010}) to the
diagnostic diagram in Figure~\ref{Luhman}. In summary, 39\,\% of the objects out of the entire
sample lie in the primordial disc region whereas between 31\,\% and 32\,\% disperse
their discs from the inside-out and up to 22\,\% of the sources have already
lost their disc. So, almost a third of the available sources are currently clearing their
discs from the inside-out. Less than 2\,\% of the objects lie in the
homogeneous draining region E. We interpret this result as strong evidence against
homogeneous disc depletion as the main disc dispersal mode.
\section{Dispersal time-scales across spectral types}
With the sample of YSOs in different star-forming regions becoming larger and
larger, it is now possible to estimate the typical time-scales for the disc dispersal
phase, even though cluster ages of course always introduce a large uncertainty in
the estimates. The disc-evolution time-scale of a star-forming region can be roughly
estimated by multiplying the age of the region by the ratio of the number of evolved
object to the total number of objects in the region. Typical transition time-scales
for the considered star-forming regions are of order $10^{5}$ yrs. The average time-scale across all spectral types is $6.9\cdot10^{5}$ yrs and roughly the same as the average cluster time-scale of $6.6\cdot10^{5}$ yrs. This is partially because there is no significant difference for timescales amongst spectral types.
We further illustrate that disc dispersal time-scales appear to be independent of
spectral type. We plot in Figure~\ref{time}, the time-scale ratio between K and M-stars and
show that this ratio is consistent with unity. This suggests that there is no significant
dependence of the time-scale on stellar mass, as has already been pointed out by
\citet{Ercolano2011}, who performed a spatial analysis of the distribution of K and
M-stars with discs in young star-forming regions and found no significant difference
in the distributions.
\begin{figure}[h]
\centering
\begin{minipage}[b]{0.5\textwidth}
\vspace{0pt}
\includegraphics[angle=90,width=\textwidth]{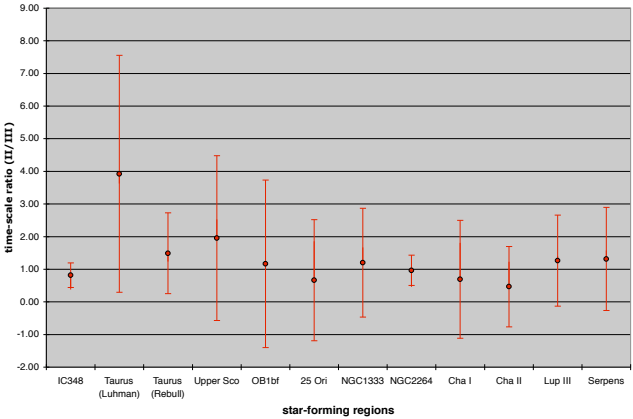} 
\end{minipage}%
\hfill
\begin{minipage}[b]{0.45\textwidth}
\vspace{0.5cm}
\caption{Time-scale ratio ${\large\tau}${\scriptsize (II)}/${\large\tau}${\scriptsize (III)} for K and M stars for the star-forming regions considered in this work with $N_{tot}>10$. Three star-froming regions are not listed, because they lack evolving objects in one or both spectral type intervals.}
\label{time}
\end{minipage}
\end{figure}
\section{Summary}
We have calculated the SEDs of protoplanetary discs of different spectral types, geometries,
settling and inclination. We then considered the evolution of the infrared
colours (K\,-\,[8] vs. K\,-\,[24]) of the model disks as they disperse according to different
scenarios (homologous depletion, inside-out and outside-in clearing). Based
on our models we propose a new diagnostic infrared colour-colour diagram to classify
the evolutionary stage of YSOs. We have applied our infrared colour-colour
diagnostic diagram to classify YSOs in 15 nearby star-forming regions and study
the evolution of their disc populations. We estimate time-scales for transition phase
of typically a few $10^{5}$ years independent of stellar mass. We conclude that, regardless
of spectral type, current observations point to fast inside-out clearing as the
preferred mode of disc dispersal.
\bibliographystyle{mn2e}
\bibliography{litnew.bib}

\begin{thebibliography}{20}
\expandafter\ifx\csname natexlab\endcsname\relax\def\natexlab#1{#1}\fi

\bibitem[{{Alexander} {et~al}\mbox{.}(2006{\natexlab{a}}){Alexander}, {Clarke},
  \& {Pringle}}]{Alexander2006a}
{Alexander} R.~D., {Clarke} C.~J., {Pringle} J.~E., 2006{\natexlab{a}}, \mnras,
  369, 216

\bibitem[{{Alexander} {et~al}\mbox{.}(2006{\natexlab{b}}){Alexander}, {Clarke},
  \& {Pringle}}]{Alexander2006b}
{Alexander} R.~D., {Clarke} C.~J., {Pringle} J.~E., 2006{\natexlab{b}}, \mnras,
  369, 229

\bibitem[{{Armitage} \& {Hansen}(1999)}]{Armitage1999}
{Armitage} P.~J., {Hansen} B.~M.~S., 1999, \nat, 402, 633

\bibitem[{{Clarke} {et~al}\mbox{.}(2001){Clarke}, {Gendrin}, \&
  {Sotomayor}}]{Clarke2001}
{Clarke} C.~J., {Gendrin} A., {Sotomayor} M., 2001, \mnras, 328, 485

\bibitem[{{Currie} \& {Kenyon}(2009)}]{Currie2009b}
{Currie} T., {Kenyon} S.~J., 2009, \aj, 138, 703

\bibitem[{{Ercolano}(2008)}]{Ercolano2008}
{Ercolano} B., 2008, in Star Formation Across the Milky Way Galaxy

\bibitem[{{Ercolano} {et~al}\mbox{.}(2011{\natexlab{a}}){Ercolano}, {Bastian},
  {Spezzi}, \& {Owen}}]{Ercolano2011}
{Ercolano} B., {Bastian} N., {Spezzi} L., {Owen} J., 2011{\natexlab{a}},
  \mnras, 416, 439

\bibitem[{{Ercolano} {et~al}\mbox{.}(2011{\natexlab{b}}){Ercolano}, {Clarke},
  \& {Hall}}]{Ercolano2010}
{Ercolano} B., {Clarke} C.~J., {Hall} A.~C., 2011{\natexlab{b}}, \mnras, 410,
  671

\bibitem[{{Ercolano} {et~al}\mbox{.}(2009){Ercolano}, {Drake}, \&
  {Clarke}}]{Ercolano2009}
{Ercolano} B., {Drake} J.~J., {Clarke} C.~J., 2009, \aap, 496, 725

\bibitem[{{Gorti} {et~al}\mbox{.}(2009){Gorti}, {Dullemond}, \&
  {Hollenbach}}]{Gorti2009}
{Gorti} U., {Dullemond} C.~P., {Hollenbach} D., 2009, \apj, 705, 1237

\bibitem[{{Haisch} {et~al}\mbox{.}(2001){Haisch}, {Lada}, \&
  {Lada}}]{Haisch2001}
{Haisch}, Jr. K.~E., {Lada} E.~A., {Lada} C.~J., 2001, \apjl, 553, L153

\bibitem[{{Hern{\'a}ndez} {et~al}\mbox{.}(2007){Hern{\'a}ndez}, {Calvet},
  {Brice{\~n}o}, {Hartmann}, {Vivas}, {Muzerolle}, {Downes}, {Allen}, \&
  {Gutermuth}}]{Hernandez2007b}
{Hern{\'a}ndez} J. {et~al.}, 2007, \apj, 671, 1784

\bibitem[{{Kenyon} \& {Hartmann}(1995)}]{Kenyon1995}
{Kenyon} S.~J., {Hartmann} L., 1995, \apjs, 101, 117

\bibitem[{{Koepferl} {et~al}\mbox{.}(2012){Koepferl}, {Ercolano}, {Dale},
  {Teixeira}, {Ratzka}, \& {Spezzi}}]{Koepferl2012}
{Koepferl} C.~M., {Ercolano} B., {Dale} J., {Teixeira} P.~S., {Ratzka} T.,
  {Spezzi} L., 2012, Disc clearing of young stellar objects: evidence for fast
  inside-out dispersal, submitted to \mnras

\bibitem[{{Luhman} {et~al}\mbox{.}(2010){Luhman}, {Allen}, {Espaillat},
  {Hartmann}, \& {Calvet}}]{Luhman2010}
{Luhman} K.~L., {Allen} P.~R., {Espaillat} C., {Hartmann} L., {Calvet} N.,
  2010, \apjs, 186, 111

\bibitem[{{Mamajek}(2009)}]{Mamajek2009}
{Mamajek} E.~E., 2009, in American Institute of Physics Conference Series, Vol.
  1158, American Institute of Physics Conference Series, {Usuda} T., {Tamura}
  M., {Ishii} M., eds., pp. 3--10

\bibitem[{{Owen} {et~al}\mbox{.}(2012){Owen}, {Clarke}, \&
  {Ercolano}}]{Owen2012a}
{Owen} J.~E., {Clarke} C.~J., {Ercolano} B., 2012, \mnras, 422, 1880

\bibitem[{{Owen} {et~al}\mbox{.}(2011{\natexlab{a}}){Owen}, {Ercolano}, \&
  {Clarke}}]{Owen2011b}
{Owen} J.~E., {Ercolano} B., {Clarke} C.~J., 2011{\natexlab{a}}, \mnras, 412,
  13

\bibitem[{{Owen} {et~al}\mbox{.}(2011{\natexlab{b}}){Owen}, {Ercolano}, \&
  {Clarke}}]{Owen2011a}
{Owen} J.~E., {Ercolano} B., {Clarke} C.~J., 2011{\natexlab{b}}, \mnras, 411,
  1104

\bibitem[{{Owen} {et~al}\mbox{.}(2010){Owen}, {Ercolano}, {Clarke}, \&
  {Alexander}}]{Owen2010}
{Owen} J.~E., {Ercolano} B., {Clarke} C.~J., {Alexander} R.~D., 2010, \mnras,
  401, 1415

\end{thebibliography}
 \end{document}